\documentclass[preprint]{ptptex}
\usepackage{wrapft}
\usepackage{graphicx}

\newcommand{\ket}[1]{| {#1} \rangle}     
\newcommand{\kket}[1]{| {#1} \rangle\!\rangle}     
\newcommand{\rbra}[1]{( {#1} |}     
\newcommand{\rket}[1]{| {#1} )}     
\newcommand{\wtilde}[1]{\widetilde{#1}} 
\newcommand{\ovl}[1]{\overline{#1}} 

\def\<{\langle}
\def\>{\rangle}
\def\bsub{\begin{subequations}}
\def\esub{\end{subequations}}
\def\beqn{\begin{eqnarray}}
\def\eeqn{\end{eqnarray}}
\def\b{\begin{equation}}

\title{
Boson Realization of the $su(3)$-Algebra. IV
}
\subtitle{
Holstein-Primakoff Representation for the Elliott Model
}    

\author{
Constan\c{c}a {\sc Provid\^encia},$^{1}$
Jo\~ao da {\sc Provid\^encia},$^{1}$\\
Yasuhiko {\sc Tsue}$^{2}$ 
and Masatoshi {\sc Yamamura}$^{3}$
}

\inst{
$^{1}$Departamento de Fisica, Universidade de Coimbra, 3004-516 Coimbra, 
Portugal\\
$^2$Physics Division, Faculty of Science, Kochi University, Kochi 780-8520, 
Japan\\
$^{3}$Faculty of Engineering, Kansai University, Suita 564-8680, Japan
}

\recdate{
\today
}

\abst{
On the basis of the Schwinger boson representation for the Elliott model 
developed in (III), the Holstein-Primakoff representation for the 
$su(3)$-algebra is presented. 
The basic idea is the same as that adopted in (II), and as an example, 
the case of the lowest approximation is shown. 
}
\begin{document}
\maketitle


This paper, (IV), is a continuation of (I),\cite{1} (II),\cite{2} 
and (III)\cite{3} and mainly we discuss the Holstein-Primakoff representation 
for the $su(3)$-algebra suitable for the Elliott model. 
The significance of the present paper may be not necessary to mention. 

In the idea similar to that presented in (II), first, we note the relation 
\bsub\label{1}
\beqn
& &({\hat a}_+^*{\hat a}_+ + {\hat a}_0^*{\hat a}_0 +{\hat a}_-^*{\hat a}_-)
\kket{Q^0,Q;L^0,LL_0}=l_a\kket{Q^0,Q;L^0,LL_0} \ , 
\label{1a}\\
& &({\hat b}_+^*{\hat b}_+ + {\hat b}_0^*{\hat b}_0 +{\hat b}_-^*{\hat b}_-)
\kket{Q^0,Q;L^0,LL_0}=l_b\kket{Q^0,Q;L^0,LL_0} \ . 
\label{1b}
\eeqn
\esub
Here, the state $\kket{Q^0,Q;L^0,LL_0}$ is defined in the relation 
(III$\cdot$5$\cdot$2) and through the relation (III$\cdot$5$\cdot$7) 
with (III$\cdot$3$\cdot$7a), we have the relation (\ref{1}), which 
gives us 
\bsub\label{2}
\beqn
& &{\hat a}_-^*{\hat a}_-\kket{Q^0,Q;L^0,LL_0}
=\left(\sqrt{l_a-{\hat a}_+^*{\hat a}_+-{\hat a}_0^*{\hat a}_0}\right)^2
\kket{Q^0,Q;L^0,LL_0}\ , 
\label{2a}\\
& &{\hat b}_-^*{\hat b}_-\kket{Q^0,Q;L^0,LL_0}
=\left(\sqrt{l_b-{\hat b}_+^*{\hat b}_+-{\hat b}_0^*{\hat b}_0}\right)^2
\kket{Q^0,Q;L^0,LL_0}\ . 
\label{2b}
\eeqn
\esub 
Of couse, the relation (\ref{2}) gives us 
\bsub\label{3}
\beqn
& &{\hat a}_-^*{\hat a}_-\kket{c(l_a,l_b)}
=\left(\sqrt{l_a-{\hat a}_+^*{\hat a}_+-{\hat a}_0^*{\hat a}_0}\right)^2
\kket{c(l_a,l_b)}\ , 
\label{3a}\\
& &{\hat b}_-^*{\hat b}_-\kket{c(l_a,l_b)}
=\left(\sqrt{l_b-{\hat b}_+^*{\hat b}_+-{\hat b}_0^*{\hat b}_0}\right)^2
\kket{c(l_a,l_b)}\ . 
\label{3b}
\eeqn
\esub 
Here, $\kket{c(l_a,l_b)}$ denotes an arbitrary superposition of the set 
$\{\kket{Q^0,Q;L^0,LL_0}\}$:
\begin{equation}\label{4}
\kket{c(l_a,l_b)}=\sum_{QLL_0}c_{QLL_0}(Q^0L^0)\kket{Q^0,Q;L^0,LL_0} \ . 
\end{equation}
Further, we should note the relation (III$\cdot$3$\cdot$8): 
\begin{equation}\label{5}
l_a=(1/2)(L^0+Q^0) \ , \qquad l_b=(1/2)(L^0-Q^0) \ . 
\end{equation}
Clearly, $({\hat a}_+, {\hat a}_+^*)$, $({\hat a}_0, {\hat a}_0^*)$, 
$({\hat b}_+, {\hat b}_+^*)$ and $({\hat b}_0, {\hat b}_0^*)$ are used 
for describing fluctuations around the equilibrium characterized by 
$(l_a,l_b)$.

Under the above argument, we can construct the Holstein-Primakoff 
representation for the Elliott model. 
For this aim, a new boson space is prepared in terms of bosons 
$({\hat \alpha}_+, {\hat \alpha}_+^*)$, 
$({\hat \alpha}_0, {\hat \alpha}_0^*)$, 
$({\hat \beta}_+, {\hat \beta}_+^*)$ and $({\hat \beta}_0, {\hat \beta}_0^*)$. 
The vacuum $\rket{0}$ of this boson space plays the same role as that 
of $\ket{T,R}$ introduced in the relation (III$\cdot$3$\cdot$2). 
In this paper, we denote $\ket{T,R}$ as $\ket{l_a,l_b}$. 
Then, we set up the following correspondence: 
\bsub\label{6}
\beqn
& &\ket{l_a,l_b} \sim \rket{0} \ , 
\label{6a}\\
& &({\hat a}_+, {\hat a}_+^*),\  ({\hat a}_0, {\hat a}_0^*), \ 
({\hat b}_+, {\hat b}_+^*)\ {\rm and}\ ({\hat b}_0, {\hat b}_0^*)\nonumber\\
& &\qquad\qquad \sim
({\hat \alpha}_+, {\hat \alpha}_+^*), \ 
({\hat \alpha}_0, {\hat \alpha}_0^*), \ 
({\hat \beta}_+, {\hat \beta}_+^*)\ {\rm and}\ 
({\hat \beta}_0, {\hat \beta}_0^*) \ , 
\label{6b}\\
& &{\hat a}_-\ {\rm and}\ {\hat a}_-^* \sim 
\sqrt{l_a-{\hat \alpha}_+^*{\hat \alpha}_+ - {\hat \alpha}_0^*{\hat \alpha}_0}
 \ , \nonumber\\
& &{\hat b}_-\ {\rm and}\ {\hat b}_-^* \sim 
\sqrt{l_b-{\hat \beta}_+^*{\hat \beta}_+ - {\hat \beta}_0^*{\hat \beta}_0} \ . 
\label{6c}
\eeqn
\esub
In the same way as that performed in (II), the relation 
(II$\cdot$2$\cdot$9) can be transcribed in the form 
\bsub\label{7}
\beqn
& &{\hat L}_+\sim {\ovl L}_+
=\sqrt{2}\Big[ {\hat \alpha}_+^*{\hat \alpha}_0+{\hat \alpha}_0^*
\sqrt{l_a-{\hat \alpha}_+^*{\hat \alpha}_+ - {\hat \alpha}_0^*{\hat \alpha}_0}
\nonumber\\
& &\qquad\qquad\qquad\qquad
+{\hat \beta}_+^*{\hat \beta}_0+
{\hat \beta}_0^*
\sqrt{l_b-{\hat \beta}_+^*{\hat \beta}_+ - {\hat \beta}_0^*{\hat \beta}_0}
\Big] \ , 
\nonumber\\
& &{\hat L}_-\sim {\ovl L}_-
=\sqrt{2}\Big[ {\hat \alpha}_0^*{\hat \alpha}_+ +
\sqrt{l_a-{\hat \alpha}_+^*{\hat \alpha}_+ - {\hat \alpha}_0^*{\hat \alpha}_0}
\ {\hat \alpha}_0 
\nonumber\\
& &\qquad\qquad\qquad\qquad
+{\hat \beta}_0^*{\hat \beta}_+ +
\sqrt{l_b-{\hat \beta}_+^*{\hat \beta}_+ - {\hat \beta}_0^*{\hat \beta}_0}\ 
{\hat \beta}_0 \Big] \ , 
\nonumber\\
& &{\hat L}_0\sim {\ovl L}_0
=2{\hat \alpha}_+^*{\hat \alpha}_+ + {\hat \alpha}_0^*{\hat \alpha}_0 
+2{\hat \beta}_+^*{\hat \beta}_+ + {\hat \beta}_0^*{\hat \beta}_0 
-(l_a + l_b) \ , 
\label{7a}\\
& &{\hat Q}_0\sim {\ovl Q}_0
=-3{\hat \alpha}_0^*{\hat \alpha}_0 
+3{\hat \beta}_0^*{\hat \beta}_0 
+(l_a - l_b) \ , 
\label{7b}\\
& &{\hat C}_1^*\sim {\ovl C}_1^*
=\sqrt{3}\Big[ {\hat \alpha}_0^*
\sqrt{l_a-{\hat \alpha}_+^*{\hat \alpha}_+ - {\hat \alpha}_0^*{\hat \alpha}_0}
-{\hat \alpha}_+^*{\hat \alpha}_0
\nonumber\\
& &\qquad\qquad\qquad\qquad
-{\hat \beta}_0^*
\sqrt{l_b-{\hat \beta}_+^*{\hat \beta}_+ - {\hat \beta}_0^*{\hat \beta}_0}
+{\hat \beta}_+^*{\hat \beta}_0 \Big] \ , 
\nonumber\\
& &{\hat C}_2^*\sim {\ovl C}_2^*
=\sqrt{6}\Big[ {\hat \alpha}_+^*
\sqrt{l_a-{\hat \alpha}_+^*{\hat \alpha}_+ - {\hat \alpha}_0^*{\hat \alpha}_0}
\nonumber\\
& &\qquad\qquad\qquad\qquad
-{\hat \beta}_+^*
\sqrt{l_b-{\hat \beta}_+^*{\hat \beta}_+ - {\hat \beta}_0^*{\hat \beta}_0}
\Big]\ , 
\nonumber\\
& &{\hat C}_1\sim {\ovl C}_1
=\sqrt{3}\Big[
\sqrt{l_a-{\hat \alpha}_+^*{\hat \alpha}_+ - {\hat \alpha}_0^*{\hat \alpha}_0}
\ {\hat \alpha}_0-{\hat \alpha}_0^*{\hat \alpha}_+
\nonumber\\
& &\qquad\qquad\qquad\qquad
-
\sqrt{l_b-{\hat \beta}_+^*{\hat \beta}_+ - {\hat \beta}_0^*{\hat \beta}_0}\ 
{\hat \beta}_0
+{\hat \beta}_0^*{\hat \beta}_+ \Big] \ , 
\nonumber\\
& &{\hat C}_2\sim {\ovl C}_2
=\sqrt{6}\Big[
\sqrt{l_a-{\hat \alpha}_+^*{\hat \alpha}_+ - {\hat \alpha}_0^*{\hat \alpha}_0}
\ {\hat \alpha}_+
\nonumber\\
& &\qquad\qquad\qquad\qquad
-
\sqrt{l_b-{\hat \beta}_+^*{\hat \beta}_+ - {\hat \beta}_0^*{\hat \beta}_0}\ 
{\hat \beta}_+ \Big] \ . 
\label{7c} 
\eeqn
\esub
The $su(1,1)$-generators and ${\wtilde R}_0$ lead us to 
\bsub\label{8}
\beqn
& &{\wtilde T}_+\sim {\check T}_+
={\hat \alpha}_0^*{\hat \beta}_0^*-{\hat \alpha}_+^*
\sqrt{l_b-{\hat \beta}_+^*{\hat \beta}_+ - {\hat \beta}_0^*{\hat \beta}_0}
-{\hat \beta}_+^*
\sqrt{l_a-{\hat \alpha}_+^*{\hat \alpha}_+ - {\hat \alpha}_0^*{\hat \alpha}_0}
\ , \nonumber\\
& &{\wtilde T}_-\sim {\check T}_-
={\hat \beta}_0{\hat \alpha}_0-
\sqrt{l_b-{\hat \beta}_+^*{\hat \beta}_+ - {\hat \beta}_0^*{\hat \beta}_0}\ 
{\hat \alpha}_+
-
\sqrt{l_a-{\hat \alpha}_+^*{\hat \alpha}_+ - {\hat \alpha}_0^*{\hat \alpha}_0}
\ {\hat \beta}_+ \ , 
\label{8a}\\
& &{\wtilde T}_0\sim {\check T}_0
=(1/2)(l_a+l_b+3) \ , 
\label{8b}\\
& &{\wtilde R}_0\sim {\check R}_0
=(1/2)(l_a-l_b) \ . 
\label{8c}
\eeqn
\esub
Thus, we arrived at our goal. 
For the construction of the physical space, we can borrow the idea used in 
(II). 
Therefore, we omit the discussion. 
We note the relation 
\begin{equation}\label{9}
{\check T}_-\rket{phys}=\rbra{phys}{\check T}_+=0 \ . 
\end{equation}

As the simplest approximation, we investigate the case where 
${\ovl L}_{\pm}$, ${\ovl C}_1^*$, ${\ovl C}_2^*$, ${\ovl C}_1$, ${\ovl C}_2$, 
${\check T}_+$ and ${\check T}_-$ are approximated in the framework of the 
linear terms for the boson operators. 
In this case, the relations (\ref{7}) and (\ref{8}) give us 
\bsub\label{10}
\beqn
& &{\ovl L}_+^{A}=\sqrt{2}(\sqrt{l_a}{\hat \alpha}_0^*
+\sqrt{l_b}{\hat \beta}_0^*) \ , \qquad
{\ovl L}_-^{A}=\sqrt{2}(\sqrt{l_a}{\hat \alpha}_0
+\sqrt{l_b}{\hat \beta}_0) \ , 
\label{10a}\\
& &{\ovl C}_1^{A*}=\sqrt{3}(\sqrt{l_a}{\hat \alpha}_0^*
-\sqrt{l_b}{\hat \beta}_0^*) \ , \qquad
{\ovl C}_2^{A*}=\sqrt{6}(\sqrt{l_a}{\hat \alpha}_+^*
-\sqrt{l_b}{\hat \beta}_+^*) \ , \nonumber\\
& &{\ovl C}_1^{A}=\sqrt{3}(\sqrt{l_a}{\hat \alpha}_0
-\sqrt{l_b}{\hat \beta}_0) \ , \qquad
{\ovl C}_2^{A}=\sqrt{6}(\sqrt{l_a}{\hat \alpha}_+
-\sqrt{l_b}{\hat \beta}_+) \ , 
\label{10b}
\eeqn
\esub
\vspace{-0.8cm}
\begin{equation}\label{11}
{\check T}_+^{A}=-(\sqrt{l_a}{\hat \alpha}_+^*
+\sqrt{l_b}{\hat \beta}_+^*) \ , \qquad
{\check T}_-^{A}=-(\sqrt{l_a}{\hat \alpha}_+
+\sqrt{l_b}{\hat \beta}_+) \ . \ \ \ 
\end{equation}
In order to rewrite the expressions (\ref{10}) and (\ref{11}), we introduce 
the following boson operators: 
\beqn
& &{\hat \gamma}_+^*=\left(\sqrt{l_a+l_b}\right)^{-1}
\left(\sqrt{l_a}{\hat \alpha}_0^*+\sqrt{l_b}{\hat \beta}_0^*\right)\ , 
\nonumber\\
& &{\hat \gamma}_1^*=\left(\sqrt{l_a+l_b}\right)^{-1}
\left(\sqrt{l_a}{\hat \alpha}_0^*-\sqrt{l_b}{\hat \beta}_0^*\right)\ , 
\nonumber\\
& &{\hat \gamma}_2^*=\left(\sqrt{l_a+l_b}\right)^{-1}
\left(\sqrt{l_a}{\hat \alpha}_+^*-\sqrt{l_b}{\hat \beta}_+^*\right)\ , 
\label{12}\\
& &{\hat \gamma}^*=\left(\sqrt{l_a+l_b}\right)^{-1}
\left(\sqrt{l_a}{\hat \alpha}_+^*+\sqrt{l_b}{\hat \beta}_+^*\right)\ , 
\label{13}
\eeqn
Then, the relations (\ref{10}) and (\ref{11}) can be rewritten as 
\bsub\label{14}
\beqn
& &{\ovl L}_+^{A}=\sqrt{2({l_a}+l_b)}{\hat \gamma}_+^* \ , \qquad
{\ovl L}_-^{A}=\sqrt{2(l_a+l_b)}{\hat \gamma}_+ \ , 
\label{14a}\\
& &{\ovl C}_1^{A*}=\sqrt{3(l_a+l_b)}\left[\frac{l_a-l_b}{l_a+l_b}
{\hat \gamma}_+^*+\frac{2\sqrt{l_al_b}}{l_a+l_b}{\hat \gamma}_1^*\right] \ , 
\nonumber\\
& &{\ovl C}_1^{A}=\sqrt{3(l_a+l_b)}\left[\frac{l_a-l_b}{l_a+l_b}
{\hat \gamma}_+ +\frac{2\sqrt{l_al_b}}{l_a+l_b}{\hat \gamma}_1\right] \ , 
\nonumber\\
& &{\ovl C}_2^{A*}=\sqrt{6(l_a+l_b)}{\hat \gamma}_2^* \ , \qquad 
{\ovl C}_2^{A}=\sqrt{6(l_a+l_b)}{\hat \gamma}_2 \ , 
\label{14b}
\eeqn
\esub
\vspace{-0.8cm}
\begin{equation}\label{15}
{\check T}_+^{A}=\sqrt{(l_a+l_b)}{\hat \gamma}^* \ , \qquad
{\check T}_-^{A}=\sqrt{(l_a+l_b)}{\hat \gamma} \ . \ \ \ \ 
\end{equation}
Further, ${\ovl L}_0$ and ${\ovl Q}_0$ shown in the relations 
(\ref{7a}) and (\ref{7b}), respectively, can be rewritten as 
\bsub\label{14-2}
\beqn
& &{\ovl L}_0=
{\hat \gamma}_+^*{\hat \gamma}_+
-\left[(l_a+l_b)-({\hat \gamma}_1^*{\hat \gamma}_1+
2{\hat \gamma}_2^*{\hat \gamma}_2)\right]+2{\hat \gamma}^*{\hat \gamma} \ , 
\label{14c}\\
& &{\ovl Q}_0=
(l_a-l_b)+3\frac{l_a-l_b}{l_a+l_b}
{\hat \gamma}_1^*{\hat \gamma}_1
-3\frac{l_a-l_b}{l_a+l_b}{\hat \gamma}_+^*{\hat \gamma}_+ 
-6\frac{\sqrt{l_al_b}}{l_a+l_b}({\hat \gamma}_1^*{\hat \gamma}_+ +
{\hat \gamma}_+^*{\hat \gamma}_1) \ , 
\label{14d}
\eeqn
\esub
The operators ${\mib C}_1^*$, ${\mib C}_2^*$ and ${\mib Q}_0$ defined in the 
relation (4$\cdot$1a), (4$\cdot$1b) and (4$\cdot$4) in (III), respectively, 
can be expressed in the following form: 
\bsub\label{16}
\beqn
& &{\mib C}_1\sim {\ovl {\mib C}}_1^{A*}=-(l_a+l_b)^3\cdot
\left(\sqrt{3(l_a+l_b)}\cdot \frac{2\sqrt{l_al_b}}{l_a+l_b}\right)\cdot
{\hat \gamma}_1^* \ , 
\label{16a}\\
& &{\mib C}_2\sim {\ovl {\mib C}}_2^{A*}=(l_a+l_b)^4\cdot
\left(\sqrt{6(l_a+l_b)}\right)\cdot
{\hat \gamma}_2^* \ , 
\label{16b}\\
& &{\mib Q}_0\sim {\ovl {\mib Q}}_0^{A}=
(l_a+l_b)^2\!\cdot\!
\left[(l_a-l_b)(1-2({\hat \gamma}_1^*{\hat \gamma}_1
-{\hat \gamma}_2^*{\hat \gamma}_2)-2{\hat \gamma}_+^*{\hat \gamma}_+)\!+\!
(l_a-l_b){\hat \gamma}^*{\hat \gamma}\right]
. \quad
\label{16c}
\eeqn
\esub
Comparison of the form (\ref{16}) with the form (\ref{14}) may be interesting. 
The operator ${\ovl C}_1^{A*}$ is expressed in terms of 
the linear combination for ${\hat \gamma}_+^*$ and ${\hat \gamma}_1^*$, 
but, ${\ovl {\mib C}}_1^{A*}$ is proportional to ${\hat \gamma}_1^*$. 
Further, in ${\ovl Q}_0$, there exists a term which express the 
coupling of ${\hat \gamma}_1$ and ${\hat \gamma}_+$ $({\hat \gamma}_1^*
{\hat \gamma}_+ + {\hat \gamma}_+^*{\hat \gamma}_1)$, but, in 
${\ovl {\mib Q}}_0^A$, there does not exist such a term.

The relations (\ref{9}) and (\ref{15}) tell us that the physical space 
is composed of the orthogonal set given by 
\begin{equation}\label{17}
\rket{m_+,m_1,m_2}=\left(\sqrt{m_+!m_1!m_2!}\right)^{-1}
({\hat \gamma}_+^*)^{m_+}({\hat \gamma}_1^*)^{m_1}({\hat \gamma}_2^*)^{m_2}
\rket{0} \ . 
\end{equation}
Let us investigate the state (\ref{17}) in detail. 
Hereafter, we omit the normalization constant. 
The state (\ref{17}) is decomposed in the form 
\begin{equation}\label{18}
\rket{m_+,m_1,m_2}=({\hat \gamma}_+^*)^{m_+}
\rket{m_1,m_2}\ , \qquad
\rket{m_1,m_2}=({\hat \gamma}_1^*)^{m_1}({\hat \gamma}_2^*)^{m_2}\rket{0} \ .
\end{equation}
Our original orthogonal set is specified by the quantum numbers 
$Q^0$, $Q$, $L^0$, $L$ and $L_0$, such as shown in the relation (\ref{1}). 
In these numbers, $Q^0$ and $L^0$ are related to $l_a$ and $l_b$ 
in the relation (\ref{5}), which specify the intrinsic state. 
Then, we investigate the relation of $m_+$, $m_1$ and $m_2$ to 
$Q$, $L$ and $L_0$, which are related to the eigenvalues of 
${\mib Q}_0$, ${\hat {\mib L}}^2$ and ${\hat L}_0$, respectively, 
for the state $\kket{Q^0,Q;L^0,LL_0}$. 
First, we note the relation 
\bsub\label{19}
\beqn
& &{\ovl {\mib Q}}_0^A\rket{m_1,m_2}=Q^A(l_a+l_b)^2\rket{m_1,m_2}\ , 
\label{19a}\\
& &Q^A=(l_a-l_b)(1-2(m_1-m_2)) \ . 
\label{19b}
\eeqn
\esub
Therefore, the eigenvalue $Q^A$, which corresponds to $Q$, is 
explicitly given in the relation (\ref{19b}). 
Next, the following relations should be noted: 
\beqn
& &{\hat \gamma}_+\rket{m_1,m_2}=0\ , \quad {\rm i.e.,}\quad
{\ovl L}_-^A\rket{m_1,m_2}=0 \ , 
\nonumber\\
& &{\ovl L}_0\rket{m_1,m_2}=-L^A\rket{m_1,m_2} \ , 
\label{20}\\
& &L^A=l_a+l_b-(m_1+2m_2)\ , 
\label{21}\\
& &\rket{m_+,m_1,m_2}=({\hat \gamma}_+^*)^{L^A+L_0^A}\rket{m_1,m_2}
=({\ovl L}_+^A)^{L^A+L_0^A}\rket{m_1,m_2} \ , 
\label{22}\\
& &L^A+L_0^A=m_+\ , \quad{\rm i.e.,}\quad 
L_0^A=m_+-(l_a+l_b-(m_1+2m_2))\ . 
\label{23}
\eeqn
Then, we can understand that $L^A$ and $L_0^A$, which correspond to 
$L$ and $L_0$, respectively, are given in the relations (\ref{21}) and 
(\ref{23}), respectively. 
The relation (\ref{20}) shows us that $L^A$ can be attributed to the 
magnitude of the angular momentum. 
Of course, this argument is valid under the condition 
\begin{equation}\label{24}
l_a+l_b \gg m_1+2m_2\ . 
\end{equation}
It is verified from the following argument: 
We define the operator ${\ovl {\mib L}}^{A2}$ in the form 
\beqn\label{25}
{\ovl {\mib L}}^{A2}&=&
{\ovl L}_0^2+(1/2)({\ovl L}_+^A{\ovl L}_-^A
+{\ovl L}_-^A{\ovl L}_+^A) \nonumber\\
&=&(l_a+l_b)(l_a+l_b+1)-4(l_a+l_b)({\hat \gamma}_2^*{\hat \gamma}_2
+{\hat \gamma}^*{\hat \gamma})-2(l_a+l_b){\hat \gamma}_1^*{\hat \gamma}_1\ . 
\eeqn
Of course, the expression (\ref{25}) is obtained in the frame of the 
bilinear for the fluctuation. 
Operation of ${\ovl {\mib L}}^{A2}$ on the state $\rket{m_+,m_1,m_2}$ 
gives us the form 
\begin{equation}\label{26}
{\ovl {\mib L}}^{A2}\rket{m_+,m_1,m_2}
=L^A(L^A+1)\rket{m_+,m_1,m_2} \ .
\end{equation}
Here, of course, we used the condition (\ref{24}). 
The above is another reason why we can attribute $L^A$ to the 
magnitude of the angular momentum. 
For the above discussion, the condition (\ref{24}) is 
very important. 
Therefore, the treatment developed in this paper may be valid 
for the case of high angular momentum. 
Finally, we comment the following point: 
The state $\rket{m_+,m_1,m_2}$ can be expressed in the form 
\begin{equation}\label{27}
\rket{m_+,m_1,m_2}=({\ovl L}_+^A)^{m_+}
({\ovl {\mib C}}_1^{A*})^{m_1}({\ovl {\mib C}}_2^{A*})^{m_2}
\rket{0} \ . 
\end{equation}
The above means that it gives us a possible form 
of the approximation for the formalism developed in (III). 
Thus, we could show that as the simplest approximation 
for the Holstein-Primakoff representation of the $su(3)$-algebra, 
we obtained an approximated form of the Elliott model, which 
may be equivalent to the RPA order.

\section*{Acknowledgements} 

On the occasion of publishing the present series of papers, one of the 
authors, M. Y., should acknowledge to Professor T. Marumori. 
Since M. Y. started study of nuclear collective motion under 
Professor T. Marumori, forty-five years have passed. 
During this period, he always suggested various ideas for physics and 
also gave encouragement to M. Y. 
The subjects discussed in this series have closed connection to 
his suggestions, for example, the RPA method, the boson mapping method, 
the algebraic aspects of many-body systems and the TDHF theory in 
canonical form.





\begin{thebibliography}{99}
\bibitem{1}
C. Provid\^encia, J. da Provid\^encia, Y. Tsue and M. Yamamura, 
submitted to Prog. Theor. Phys. ; nucl-th/.
\bibitem{2}
C. Provid\^encia, J. da Provid\^encia, Y. Tsue and M. Yamamura, 
submitted to Prog. Theor. Phys. ; nucl-th/.
\bibitem{3}
C. Provid\^encia, J. da Provid\^encia, Y. Tsue and M. Yamamura, 
submitted to Prog. Theor. Phys. ; nucl-th/.
\end{thebibliography}
\end{document}